\documentclass[12pt]{article}
\usepackage{amsmath,amsfonts}
\usepackage{graphicx}
\input epsf

\makeatletter\@addtoreset{equation}{section}\makeatother

\def\be{\begin{equation}}
\def\ee{\end{equation}}
\def\bea{\begin{eqnarray}}
\def\eea{\end{eqnarray}}

\makeatletter\@addtoreset{equation}{section}\makeatother

\renewcommand{\title}[1]{\vbox{\center\LARGE{#1}}\vspace{5mm}}
\renewcommand{\author}[1]{\vbox{\center#1}\vspace{5mm}}
\newcommand{\address}[1]{\vbox{\center\em#1}}
\newcommand{\email}[1]{\vbox{\center\tt#1}\vspace{5mm}}

\begin{document}

\unitlength = .8mm

\begin{titlepage}
\begin{center}
\hfill \\
\hfill \\
\vskip 1cm

\title{Chern-Simons-Matter Theory and Mirror Symmetry}

\author{Daniel Louis Jafferis$^{1,a}$,
Xi Yin$^{2,b}$}

\address{$^1$NHETC and Department of Physics and Astronomy, Rutgers University,\\ Piscataway, NJ 08855, USA\\
\medskip
$^2$Jefferson Physical Laboratory, Harvard University,\\
Cambridge, MA 02138 USA}

\email{$^a$jafferis@physics.rutgers.edu,
$^b$xiyin@fas.harvard.edu}

\end{center}

\abstract{ In this paper we study supersymmetric
Chern-Simons-matter (CSM) theories with several Higgs
branches. Two such theories at small Chern-Simons level
are conjectured to describe the superconformal field theory at the
infrared fixed point of ${\cal N}=4$ QED with $N_f=2,3$. In
particular, the mirror symmetry which exchanges the Coulomb and
Higgs branches of ${\cal N}=4$ QED with $N_f=2$ is manifest in the
Chern-Simons-matter description. We also study the quantum corrections
to the moduli space of a class of ${\cal N}=2$ CSM theories. }

\vfill

\end{titlepage}

\eject \tableofcontents

\section{Introduction}

Chern-Simons-matter (CSM) theories in 2+1 dimensions provide a
large class of \\ (super-)conformal field theories with Lagrangian
descriptions \cite{Ivanov:1991fn, Avdeev:1991za, Avdeev:1992jt,
Kapustin:1999ha, Schwarz:2004yj, Gaiotto:2007qi, Gaiotto:2008sd,
Hosomichi:2008jd}. These theories have received much attention
recently due to the discovery of their $AdS_4$ gravity duals
\cite{Aharony:2008ug} (see also \cite{Benna:2008zy,
Hosomichi:2008jb, Aharony:2008gk, Imamura:2008nn, Jafferis:2008qz,
Martelli:2008si}). In this paper, we explore abelian CSM theories
whose moduli space has different Higgs branches that meet at a
point, exhibiting the quantum criticality previously well known
between the Coulomb and Higgs branch of 2+1 dimensional gauge
theories \cite{Seiberg:1996nz,
Intriligator:1996ex,Aharony:1997bx}.

We start with a class of ${\cal N}=4$ CSM theories with $U(1)\times U(1)$ gauge group at level
$(k,-k)$,
which are of the BF type studied in
\cite{Kapustin:1999ha}. These theories have two Higgs branches, parameterized
by the scalars of the hypermultiplet matter fields.
An important ingredient is that the hypermultiplet moduli spaces
in CSM theories can receive quantum corrections, unlike in 2+1 dimensional
${\cal N}=4$ Yang-Mills theories. The quantum corrected Higgs branches of these theories
turn out to be the same as the Coulomb and Higgs branches of the infrared superconformal fixed point of
${\cal N}=4$ $U(1)$ gauge theory with $N_f$ charged hypermultiplets.
This leads us to conjecture that the CSM theory describes the {\sl same} SCFT as the
IR limit of ${\cal N}=4$ SQED!

In the cases $N_f=2,3$,  we give further evidence for this
equivalence using a brane construction, and find that the infrared
theory describes a certain fractional M2 brane in an ${\cal N}=4$
orbifold. The ${\cal N}=4$ SQED with $N_f=2$ is known
\cite{Intriligator:1996ex} to be self-mirror and has enhanced
global symmetry at the IR fixed point. The mirror symmetry
exchanges the Coulomb and Higgs branches of the theory. In the
dual CSM description, the mirror symmetry is manifest in the
Lagrangian, and exchanges the two Higgs branches. The enhanced
global symmetry in the CSM theory can be understood in terms of 't
Hooft operators, which allows the construction of new symmetry
currents at small Chern-Simons levels.

We then move to ${\cal N}=2$ CSM theories and ask whether they can
have different Higgs branches meeting at a quantum critical point.
This is easy to realize in the classical theory. Quantum
mechanically, the moduli spaces receive nontrivial corrections. In
fact, in the class of ${\cal N}=2$ theories we will consider, the
classical moduli space will always be lifted by quantum effects
due to the shift of Chern-Simons level when massive charged chiral
multiplets are integrated out. This problem can be avoided if we
start with shifted bared CS levels, such that the classical moduli
space is lifted by the D-term potential, but the moduli space is
restored when quantum corrections are taken into account. We find
that the one-loop corrected metric of the Higgs branch is the
K\"ahler metric of a symplectic quotient space of the form
${\mathbb C}^{M+1}//U(1)$.

The paper is organized as follows. Section 2  describes the Lagrangian of a class of ${\cal N}=4$ CSM
theories. The moduli spaces of these theories are studied in section 3.
In section 4 we describe the brane constructions, and argue that these theories
describe the same SCFT as that of the IR fixed point of ${\cal N}=4$ QED with $N_f$ flavors. We also discuss a simple nonabelian generalization.
In section 5, we study the moduli space of a class of ${\cal N}=2$ CSM theories with no superpotential.
We conclude in section 6.

\section{${\cal N}=4$ Chern-Simons-matter theories as quantum critical points}

\subsection{Model II}

We start by considering ${\cal N}=3$ Chern-Simons-matter theory
with gauge group $U(1)\times U(1)$, at Chern-Simons level
$(k,-k)$, and matter hypermultiplets
$(X_i,\tilde X_i)$ of charge $(+1,+1)$ and $(-1,-1)$,
$i=1,\cdots,N_1$, and $(Y_{i'},\tilde Y_{i'})$ of charge $(+1,-1)$
and $(-1,+1)$, $i'=1,\cdots,N_2$. In ${\cal N}=2$ language, we have gauge multiplets
$(A_\mu,\sigma,D;\chi)$ and $(\tilde A_\mu, \tilde\sigma, \tilde
D; \tilde\chi)$,
and chiral multiplets $X_i, \tilde X_i, Y_{i'},
\tilde Y_{i'}$.
The $D$-term scalar potential before integrating out the auxiliary fields is
\begin{equation}
\begin{aligned}
 V_D&={k\over2\pi} D\sigma - {k\over2\pi} \tilde D\tilde\sigma + \sum_i
(\sigma+\tilde\sigma)^2(|X_i|^2+|\tilde X_i|^2) +
\sum_{i'} (\sigma-\tilde\sigma)^2(|Y_{i'}|^2+|\tilde Y_{i'}|^2) \\
&~+ D(\sum_i |X_i|^2-\sum_i|\tilde X_i|^2+\sum_{i'}|Y_{i'}|^2-\sum_{i'} |\tilde Y_{i'}|^2) \\
&~ +\tilde D (\sum_i |X_i|^2-\sum_i|\tilde X_i|^2-\sum_{i'} |Y_{i'}|^2+\sum_{i'} |\tilde Y_{i'}|^2).
\end{aligned}
\end{equation}
Integrating out $D$ and $\tilde D$ sets
\begin{equation}
\begin{aligned}
& \sigma = -{2\pi\over k} (\sum_i |X_i|^2-\sum_i|\tilde X_i|^2+\sum_{i'}|Y_{i'}|^2-\sum_{i'} |\tilde Y_{i'}|^2), \\
&\tilde\sigma = {2\pi\over k} (\sum_i |X_i|^2-\sum_i|\tilde X_i|^2-\sum_{i'} |Y_{i'}|^2+\sum_{i'} |\tilde Y_{i'}|^2).
\end{aligned}
\end{equation}
So we obtain the scalar potential
\begin{equation}
\begin{aligned}
V_D &= {16\pi^2\over k^2} \left[ \left(\sum_i |X_i|^2-\sum_i|\tilde X_i|^2\right)^2 \left(\sum_{i'} |Y_{i'}|^2
+\sum_{i'}|\tilde Y_{i'}|^2\right) \right.\\
&\left.~~~+\left(\sum_{i'} |Y_{i'}|^2-\sum_{i'}|\tilde Y_{i'}|^2\right)^2 \left(\sum_{i} |X_{i}|^2
+\sum_i |\tilde X_i|^2\right) \right]
\end{aligned}
\end{equation}
There is an ${\cal N}=2$ superpotential
\begin{equation}
W = {8\pi\over k} \sum_iX_i\tilde X_i\sum_{j'}Y_{j'}\tilde Y_{j'}
\end{equation}
giving rise to the $F$-term scalar potential
\begin{equation}
V_F = {64\pi^2\over k^2} \left[ \left|\sum_i X_i \tilde X_i\right|^2
\left(\sum_{i'} |Y_{i'}|^2
+\sum_{i'}|\tilde Y_{i'}|^2\right)
+ \left|\sum_{i'} Y_{i'} \tilde Y_{i'}\right|^2 \left(\sum_{i} |X_{i}|^2
+\sum_i |\tilde X_i|^2\right) \right]
\end{equation}
The Higgs branch moduli space contains the locus where all $X_i$'s vanish and $Y_{i'}$'s arbitrary, and
all $Y_{i'}$'s vanish and $X_i$'s arbitrary. However, for $N_1, N_2>1$, there is also the locus given by
\begin{equation}
\begin{aligned}
&\sum |X_i|^2-\sum |\tilde X_i|^2 =0,~~~\sum X_i\tilde X_i=0,\\
&\sum |Y_{i'}|^2-\sum |\tilde Y_{i'}|^2 =0,~~~\sum Y_{i'}\tilde Y_{i'}=0.
\end{aligned}
\end{equation}
so that the entire Higgs moduli space is connected. A special case, however, is when
$N_1=N_2=1$, where the scalar potential becomes
\begin{equation}\label{Vpot}
V = {16\pi^2\over k^2} \left[ (|X|^2+|\tilde X|^2)^2 (|Y|^2+|\tilde Y|^2)
+ (|Y|^2+|\tilde Y|^2)^2 (|X|^2+|\tilde X|^2)\right]
\end{equation}
In manifestly $SU(2)$ R-symmetry invariant notation, we can write
$X_a=(X, \bar {\tilde X})$, $Y_a = (Y,\bar{\tilde Y})$, $\xi^a = (\xi,\bar{\tilde\xi})$, $\eta^a=(\eta,\bar{\tilde \eta})$. The fermion-boson coupling in the ${\cal N}=3$
CSM with $N_1=N_2=1$ is given by
\begin{equation}\label{LF}
\begin{aligned}
L_F &= {4\pi\over k}\left[
(|X|^2-|\tilde X|^2) (\bar\eta\eta-\bar{\tilde\eta}\tilde\eta)+
(|Y|^2-|\tilde Y|^2)(\bar\xi\xi-\bar{\tilde \xi}\tilde \xi) \right]
+{8\pi\over k} \left[(X\bar \xi-\tilde X\bar{\tilde\xi})(\bar Y\eta-\bar{\tilde Y}\tilde\eta)+c.c\right]\\
&~~~+{8\pi\over k} \left(X\tilde X\eta\tilde\eta + Y\tilde Y\xi\tilde \xi + XY\tilde\xi\tilde \eta + \tilde X\tilde Y \xi\eta
+X\tilde Y \tilde \xi\eta + \tilde X Y \xi\tilde\eta + c.c. \right)
\\
&= {8\pi\over k} \left( \bar X_{(a} X_{b)} \bar \eta^{(a} \eta^{b)}
+  \bar Y_{(a} Y_{b)} \bar \xi^{(a} \xi^{b)}
+ \bar X_a \bar\eta^a Y_b \xi^b +X_a \eta^a \bar Y_b \bar \xi^b
+X_a \bar \eta^a Y_b \bar\xi^b + \bar X_a \eta^a \bar Y_b \xi^b
\right)
\end{aligned}
\end{equation}
Now we can see that the theory in fact has $SU(2)_L\times SU(2)_R$
symmetry, under which $X$ and $\eta$ transform as $(2,1)$ whereas
$Y$ and $\xi$ transform as $(1,2)$. The ${\cal N}=3$ $SU(2)$
R-symmetry is the diagonal subgroup of $SU(2)_L\times SU(2)_R$.
Hence we see that the ${\cal N}=3$ supersymmetry is in fact
enhanced to ${\cal N}=4$. $X$ and $Y$ then become ${\cal N}=4$ hypermultiplet
and twisted hypermultiplet, respectively. We shall refer to this theory at level $(k,-k)$ as ``Model
II${}_k$". It is in fact the same as the ${\cal N}=4$ BF theory studied in \cite{Kapustin:1999ha}. In the $SU(2)_L\times SU(2)_R$ invariant notation, we can
write the fermion-boson coupling as
\begin{equation}\label{LFtwo}
\begin{aligned}
L_F &= {8\pi\over k} \left[ \bar X_{(a} X_{b)} \bar \eta^{(a} \eta^{b)}
+  \bar Y_{(A} Y_{B)} \bar \xi^{(A} \xi^{B)}
+ \bar X_a \bar\eta^a Y_A \xi^A + \bar Y_A \bar \xi^A X_a \eta^a
+X_a \bar \eta^a Y_A \bar\xi^A + \bar Y_A \bar\xi^A \bar X_a \eta^a
\right]
\end{aligned}
\end{equation}

A slightly more general case is the ${\cal N}=4$ $U(1)_k\times
U(1)_{-k}$ CSM theory with $N_1=N_f-1$, $N_2=1$. We shall refer to
this theory as model II$(N_f)_k$. It has scalar potential
\begin{equation}
\begin{aligned}
V &= {16\pi^2\over k}  \left(|Y|^2+|\tilde Y|^2\right)^2 \left(\sum_i |X_i|^2+\sum_i |\tilde X_i|^2\right)
\\
&+{16\pi^2\over k} \left[  \left(\sum_i |X_i|^2-\sum_i |\tilde X_i|^2\right)^2
+4\left|\sum_i X_i\tilde X_i\right|^2 \right]\left(|Y|^2+|\tilde Y|^2\right)
\end{aligned}
\end{equation}
There are two branches of Higgs moduli spaces, ${\cal M}_X$ of
complex dimension $2(N_f-1)$ parameterized by arbitrary
$X_i,\tilde X_i$ and vanishing $Y,\tilde Y$, and ${\cal M}_Y$ of
complex dimension 2 parameterized by arbitrary $Y,\tilde Y$ and
vanishing $X_i,\tilde X_i$. They meet at the origin. Due to the
Mukhi effect \cite{Mukhi:2008ux, Aharony:2008ug}, the moduli space
is modded out by a discrete group of constant gauge
transformations. We will discuss this as well as the quantum
corrections to the moduli space in section 3. For now, we note
that the $X$-branch of the moduli space, ${\cal M}_X$, should be
singular along the locus
\begin{equation}\label{sing}
{\cal S}_X:~~~\sum_i |X_i|^2-\sum_i |\tilde X_i|^2 = \sum_i X_i\tilde X_i=0,
\end{equation}
where $Y$ and $\tilde Y$ becomes massless.

\subsection{Model III}

The next model we shall consider is the ${\cal N}=3$
$U(1)_k\times U(1)_{-k}\times U(1)_k$ CSM theory, with
hypermultiplet $(X,\tilde X)$ of charges $(+1,-1,0)$ and
$(-1,+1,0)$ and $(Y,\tilde Y)$ of charge $(0,-1,+1)$ and
$(0,+1,-1)$. We will refer to this theory as ``model III". The
overall $U(1)$ in this theory decouples, and can be integrated
out. Denote the gauge fields of the three $U(1)$'s by $A_1, A_2,
A_3$, and define $a=A_1-A_2$, $b=A_3-A_2$, then the CS term is
\begin{equation}
\begin{aligned}
&{k\over 4\pi} \int (A_1 \wedge dA_1 -A_2\wedge dA_2+A_3\wedge dA_3) \\
&= {k\over 4\pi} \int \left[(A_2+a)\wedge d(A_2+a) -A_2 \wedge dA_2 + (A_2+b)\wedge d(A_2+b)\right]
\end{aligned}
\end{equation}
Now $A_2$ decouples from the matter fields, and up to a gauge transformation its equation of motion sets
$A_2=-a-b$, and the CS term becomes
\begin{equation}
-{2k\over 4\pi} \int a\wedge db
\end{equation}
Upon redefining $a_\mu=A_\mu+\tilde A_\mu$, $b_\mu=A_\mu-\tilde
A_\mu$, we recover model II at level $2k$. Therefore, we see that
III${}_{k}$ is the same theory as II${}_{2k}$.

\subsection{Model IV}

Now consider the ${\cal N}=3$ $U(1)_{-k}\times U(1)_k\times U(1)_{-k}\times U(1)_k$ CSM theory, with hypermultiplets $(X,\tilde X)$, $(Y,\tilde Y)$, $(Z,\tilde Z)$, of charges
$(+1,-1,0,0)$, $(0,+1,-1,0)$ and $(0,0,+1,-1)$. This theory will be denoted model IV${}_k$. It has in fact
${\cal N}=4$ supersymmetry as well. Writing the CS action as
\begin{equation}
S_{CS} = {k\over 4\pi}\int (-A_1\wedge dA_1+A_2\wedge dA_2 - A_3 \wedge dA_3+A_4\wedge dA_4),
\end{equation}
the overall $U(1)$ decouples from the matter fields. Writing
$a=A_1-A_2$, $b=A_2-A_3$, $c=A_3-A_4$, then integrating out $A_4$
sets $a+c=0$. The CS action now reduces to
\begin{equation}
{k\over 2\pi} \int a\wedge db
\end{equation}
and the hypermultiplets $X,Y,Z$ have charges $(1,0)$, $(0,1)$ and $(-1,0)$ under $(a,b)$. Redefining $a=A+\tilde A$
and $b=A-\tilde A$, the fields $X,Y,Z$ now have charges $(+1,+1)$, $(+1,-1)$ and $(-1,-1)$ under $(A,\tilde A)$. We can interchange $Z$ with $\tilde Z$, and hence model IV${}_k$ is the same as model II$(N_f=3)_{2k}$ introduced earlier.

\section{Quantum corrections to hypermultiplet moduli space}

In 2+1 dimensional ${\cal N}=4$ Yang-Mills theories coupled to hypermultiplet matter fields, the Higgs branch moduli space,
i.e. the moduli space of hypermultiplets, is not corrected by quantum effects because the Yang-Mills coupling constant
can be promoted to a vector superfield, which decouples from the hypermultiplets at the level of kinetic terms. This
non-renormalization argument does not apply to ${\cal N}=4$ CSM theories \cite{Gaiotto:2007qi}. So the hypermultiplet moduli space
in general can and in fact will get quantum corrections, as we shall argue.

Let us start with model II$(N_f)_k$. There are $N_f-1$ hypermultiplets $X_i$ of charge $(+1,+1)$ and 1 hypermultiplet $Y$ of charge
$(+1,-1)$, under the $U(1)\times U(1)$ gauge fields $A_\mu, \tilde A_\mu$.Write $a_\mu=A_\mu+\tilde A_\mu$, $b_\mu=A_\mu-\tilde A_\mu$, and let $n_a$ and $n_b$ be the magnetic flux
of $a_\mu$ and $b_\mu$ on the sphere at infinity. The constant gauge transformations $e^{2\pi i \eta_a}$ and $e^{2\pi i \eta_b}$ satisfy
\begin{equation}
{k\over 2}(\eta_a n_b+\eta_b n_a) \in {\mathbb Z}
\end{equation}
where the factor of $1/2$ is due to the normalization of the twisted CS term $\int a \wedge db$. A priori,
it follows from Dirac quantization condition that $n_a$ and $n_b$ should be integer valued. So the constant gauge transformations are given by
\begin{equation}
\eta_a,\eta_b \in {2\over k}{\mathbb Z},
\end{equation}
For even $k$, they generate the subgroup ${\mathbb Z}_{k/2}\times {\mathbb Z}_{k/2}\subset U(1)\times U(1)$. In the previous section
we have obtained model II$(N_f)_k$ with even $k$ by integrating out the overall $U(1)$ in model III and IV, the latter involving only
bifundamentals and can be embedded straightforwardly in string theory. So we will assume $k$ is even in model II for now.

The classical moduli space has two Higgs branches ${\cal M}_X$ and ${\cal M}_Y$. After modding out by constant gauge symmetries, we have
${\cal M}_X^{cl}\simeq {\mathbb C}^{2(N_f-1)}/{\mathbb Z}_{k/2}$, and ${\cal M}_Y^{cl}\simeq {\mathbb C}^2/{\mathbb Z}_{k/2}$.
Let us first consider ${\cal M}_Y^{cl}$. Along this moduli space, the fields $(X_i,\tilde X_i)$ are massive, and can be integrated out.
The hypermultiplet mass for $(X_i,\tilde X_i)$ is a triplet of an $SU(2)$ R-symmetry,
\begin{equation}
{4\pi\over k} \bar Y_{(a} Y_{b)} = {4\pi\over k} \left({1\over 2}(|Y|^2-|\tilde Y|^2), Y\tilde Y, \bar Y\bar{\tilde Y}\right)
\equiv \vec{m}_X
\end{equation}
${\cal M}_Y^{cl}$ is a $S^1$-bundle over the ${\mathbb R}^3$ parameterized by $\vec m_X$, such that on any $S^2$ around the origin of ${\mathbb R}^3$
the fibration is a circle bundle of degree $k/2$. We propose that the effect of integrating out {\sl one} hypermultiplet of
mass $\vec m_X$ is to shift the degree of this circle bundle by $+1$. The topology of ${\cal M}_Y$ together with the rigidity of
hyperk\"ahler metrics and the homogeneity in $(Y,\tilde Y)$ then fixes ${\cal M}_Y$. After integrating out $N_f-1$ hypermultiplets, we end up with
${\cal M}_Y \simeq {\mathbb C}^2/{\mathbb Z}_{{k\over 2}+N_f-1}$.

To justify our proposal, one may compute the correction to the metric on ${\cal M}_Y$ by integrating out $(X_i,\tilde X_i)$ at one-loop.
As well known in the Coulomb branch moduli space in SQED \cite{Seiberg:1996nz, Intriligator:1996ex, Sachdev},
the one-loop contribution to the kinetic term from a hypermultiplet of mass $\vec m$ coupled to gauge field $a_\mu$ takes the form
\begin{equation}
\begin{aligned}
&\int {1\over 8\pi|\vec m|} (\partial_\mu\vec m\cdot \partial^\mu\vec m- |d a|^2)+ \epsilon^{\mu\nu\rho}\epsilon^{ijk}\partial_i({1\over 8\pi|\vec m|}) a_\mu \partial_\nu
m_j \partial_\rho m_k \\
&= \int {1\over 8\pi|\vec m|} (\partial_\mu\vec m\cdot \partial^\mu\vec m- |d a|^2)+ (*da)^\mu \omega_i \partial_\mu m_i
\end{aligned}
\end{equation}
where $\omega_i$ is the vector potential of a Dirac monopole in
the $\vec m$-space. We can dualize $a_\mu$ by replacing $da$ with
an independent two-form field $\tilde F_a$ and introducing the
Lagrangian multiplier field $\varphi_Y$ of periodicity 1. The
bosonic part of the action is
\begin{equation}
\begin{aligned}
& \int |D_\mu Y|^2+{1\over 8\pi|\vec m|} (\partial_\mu\vec m\cdot \partial^\mu\vec m- |\tilde F_a|^2)+
\int \tilde F_a\wedge (d\varphi_Y+\omega_i dm_i+{k\over 4\pi}b)
\end{aligned}
\end{equation}
Integrating out $\tilde F_a$ gives
\begin{equation}
\begin{aligned}
& \int |D_\mu Y|^2+{1\over 8\pi|\vec m|} \partial_\mu\vec m\cdot \partial^\mu\vec m+2\pi |\vec m|
(\partial_\mu\varphi_Y+\omega_i \partial_\mu m_i+{k\over 4\pi}b_\mu)^2
\end{aligned}
\end{equation}
Finally, integrating out $b_\mu$ then identifies ${2\over
k}\varphi_Y$ with the overall phase of $Y$. The $U(1)$ gauge
symmetry acts on the fields as
\begin{equation}
Y\to e^{i\Lambda}Y,~~~b_\mu\to b_\mu+\partial_\mu \Lambda,~~~\varphi_Y\to \varphi_Y -{k\over 4\pi}\Lambda
\end{equation}
The constant gauge transformations ${\mathbb Z}_{k/2}$ act on $Y$
as $Y\to e^{4\pi i/k}Y$, i.e. the phase of $Y$ has periodicity
$4\pi/k$. Therefore the effect of the one-loop correction is to
have the $S^1$ parameterized by the phase of $Y$ fibered over
$\mathbb{R}^3=\{\vec m\}$ with degree shifted from ${k\over 2}$ to
${k\over 2}+1$. This is due to the coupling $k |{\vec m}| b^\mu
\omega_i \partial_\mu m_i$, which effectively shifts phase
rotation of $Y$ when one sends $\varphi_Y \rightarrow \varphi_Y +
1$. Similarly, when $N_f-1$ $(X_i,\tilde X_i)$'s are integrated
out, the degree is shifted from ${k\over 2}$ to ${k\over
2}+N_f-1$.

Let us now consider the moduli space ${\cal M}_X$. The hypermultiplet mass of $(Y,\tilde Y)$ along ${\cal M}_X^{cl}$ is given by
\begin{equation}\label{mY}
{4\pi\over k} \sum_{i=1}^{N_f-1}\bar X^i_{(a} X_{ib)} = {4\pi\over k} \left({1\over 2}\sum_i(|X_i|^2-|\tilde X^i|^2), \sum_iX_i\tilde X^i, \sum_i\bar X_i\bar{\tilde X}^i\right)
\equiv \vec{m}_Y
\end{equation}
We can think of (\ref{mY}) as a fibration of ${\cal M}_Y^{cl}$ over ${\mathbb R}^3=\{\vec m_Y\}$, whose fiber
is an $S^1$-bundle $L_{k/2}$ over $T^*\mathbb{CP}^{N_f-2}$. Here $L_{k/2}$ is fibered over the $\mathbb{CP}^{N_f-2}$ with degree $k/2$, due to quotienting by constant gauge symmetries ${\mathbb Z}_{k/2}$.
Following our discussion above, the effect of integrating out
$(Y,\tilde Y)$ is to tensor this $S^1$-bundle (which is also fibered over ${\mathbb R}^3$) with the degree $+1$ circle bundle over ${\mathbb R}^3-\{0\}$. Spelling this out explicitly, the corrected moduli space ${\cal M}_Y$ can be expressed as
\begin{equation}\label{mYY}
\begin{aligned}
&\left({1\over 2}\sum_i(|X_i|^2-|\tilde X^i|^2), \sum_iX_i\tilde X^i, \sum_i\bar X_i\bar{\tilde X}^i\right)
= {k\over 4\pi}\vec{m}_Y,\\
&\left({1\over 2}(|Q|^2-|\tilde Q|^2), Q\tilde Q, \bar Q\bar{\tilde Q}\right)
= {k\over 4\pi}\vec{m}_Y,\\
&{\rm modulo}~~U(1):~(X_i\to e^{{2i\over k}\theta} X_i,~ \tilde X_i\to e^{-{2i\over k}\theta}\tilde X_i,~ Q\to e^{-i\theta}Q,~\tilde Q\to e^{i\theta}\tilde Q),
\end{aligned}
\end{equation}
where we have introduce the variables $(Q,\tilde Q)$ to parameterize an $S^1$ fibered over ${\mathbb R}^3-\{0\}$ of degree $1$,
so that after quotienting by the $U(1)$, the degree of $L_{k/2}$ over ${\mathbb R}^3-\{0\}$ is shifted by $+1$.
In other words, ${\cal M}_Y$ is the hyperk\"ahler quotient of ${\mathbb C}^{2N_f}////U(1)$, where the coordinates $(X_i, \tilde X_i, Q,\tilde Q)$ are assigned charges $(1,-1,-{k\over2},{k\over2})$ under the $U(1)$. For $N_f=2$, ${\cal M}_Y$ reduces to ${\mathbb C}^2/{\mathbb Z}_{{k\over 2}+1}$.

\section{Brane construction, enhanced global symmetry, and ${\cal N}=4$ QED}

\subsection{Model III and ${\cal N}=4$ QED with $N_f=2$}

Our model III${}_k$ can be engineered in type IIB string theory
\cite{Hanany:1996ie, Kitao:1998mf, Bergman:1999na, Gaiotto:2008sd}
as a D3-brane suspended from one NS5-brane, across a $(1,k)$
5-brane and a NS5-brane, to another $(1,k)$ 5-brane, arranged to
preserve ${\cal N}=3$ supersymmetry. After putting this system on
a circle, taking the small radius limit, performing T-duality and
lifting to M-theory, we end up the toric hyperk\"ahler manifold
$X_8$:
\begin{equation}
\begin{aligned}
&ds_{X_8}^2 = U_{ab}d\vec x^a \cdot d\vec x^b + U^{ab} (d\phi_a+A_a)(d\phi_b+A_b), \\
& U_{ab} = {2\over |\vec x_1|}\left(\begin{array}{cc} 1 & 0 \\ 0 & 0 \end{array} \right)
+{2\over |\vec x_1+k \vec x_2|}\left(\begin{array}{cc} 1 & k \\ k & k^2 \end{array} \right)
\end{aligned}
\end{equation} The suspended D3 brane dualizes to a fractional M2
brane at the singularity at the origin in this space, since the D3
did not entirely wrap the circle. After the change of variables
\begin{equation}
\vec x_1' = \vec x_1,~~~\vec x_2'=\vec x_1+k\vec x_2,~~~\phi_1' = \phi_1-{1\over k}\phi_2, ~~~\phi_2'={1\over k}\phi_2.
\end{equation}
we see that $X_8$ is $({\mathbb C}^2/{\mathbb Z}_2)^2/{\mathbb
Z}_k$. For general $k>2$, $X_8$ has symmetry $SU(2)\times SU(2)$.
This is the R-symmetry of the ${\cal N}=4$ CSM theory. For
$k=1,2$, however, the symmetry of $X_8$ is enhanced to
$SO(4)\times SO(4)\simeq SU(2)_F\times SU(2)_F'\times
SU(2)_L\times SU(2)_R$. The D3-brane has turned into a fractional
M2-brane sitting at the singularity of $X_8$.

Note that model III
at level $k=1$ has the same moduli space and global symmetry as
the IR SCFT of ${\cal N}=4$ $U(1)$ gauge theory with $N_f=2$
hypermultiplet matter fields. We shall refer to this SCFT as
SQED-2. The quantum moduli space of both theories are
two branches of ${\mathbb C}^2/{\mathbb Z}_2$, meeting at the origin.
We shall argue that III${}_1$ and SQED-2 are in fact the
same SCFT.

We can start with the brane configuration of a pair of NS5-branes
and a pair of D5-branes in between, and a D3-brane stretching from
one NS5 to the other, crossing the two D5-branes. Explicitly, the
configuration is (with the axion $\chi$ set to zero)

\begin{tabular}{ccccccccccc}
& 0& 1& 2& 3& 4& 5& 6& 7& 8& 9 \\
NS5& $\times$& $\times$& $\times$& 0& $\times$ & $\times$& $\times$ & & &\\
NS5$'$& $\times$&$\times$ & $\times$ & $L$& $\times$ & $\times$& $\times$ & & &\\
D5 & $\times$& $\times$ & $\times$ & $y_1$&  & & &$\times$ & $\times$& $\times$ \\
D5$'$ & $\times$& $\times$ & $\times$ & $y_2$&  & & &$\times$ & $\times$& $\times$ \\
D3 & $\times$ & $\times$ & $\times$ & $\times$ &
\end{tabular}

\noindent with $0<y_1<y_2<L$. The low energy world volume theory
on the D3-brane is 2+1 dimensional ${\cal N}=4$ QED with $N_f=2$.
Now we can move D5$'$ to the right, crossing NS5$'$, and end up
with $0<y_1<L<y_2$. A single D3 brane is created between the
NS5$'$ and the D5$'$ by the Hanany-Witten effect
\cite{Hanany:1996ie}, so the D3-brane is now stretched from NS5 to
D5$'$. We can perform the $SL(2,{\mathbb Z})$ duality $\tau\to
\tau/(1-\tau)$, so that the D5-branes are turned into $(1,1)$
5-branes, while the NS5-branes stay the same. The low energy world
volume theory is now described by model III at level $k=1$. This
leads us to the conjecture that model III${}_1$ is the same as
SQED-2.

The SQED-2 has been conjectured to have enhanced $SU(2)_F\times
SU(2)_F'\times SU(2)_L\times SU(2)$ global symmetry, where the
$SU(2)_F'$ is not manifest in the UV description, and emerges at
the IR fixed point \cite{Intriligator:1996ex}. The $SU(2)_F$ and $SU(2)_F'$ are exchanged by
mirror symmetry. In model III${}_{k=1}$, neither $SU(2)_F$ nor
$SU(2)_F'$ is manifest in the Lagrangian of the theory, but the
mirror symmetry is a manifest ${\mathbb Z}_2$ symmetry of the
Lagrangian, and so is the symmetry between the two branches of the
Higgs moduli space.

To see how the $SU(2)_F\times SU(2)_F'$ emerges in model III at $k=1$, or model II at $k=2$, note firstly that for II${}_{k=1,2}$, by combining with a 't Hooft
operator, we can shift the $U(1)$ charge of $\bar X_a$ to that of $X_a$, etc. We shall denote the field $\bar X_a$
dressed by the 't Hooft operator as $C\bar X_a$. This makes it possible for
$X_a$ and $C \bar X_a$ to fit into a multiplet of $SU(2)_F\times SU(2)_L$, for some $SU(2)_F$ symmetry.

So we may group the fields into multiplets
of $SU(2)_F\times SU(2)_F'\times SU(2)_L\times SU(2)_R$ for some $SU(2)_F\times SU(2)_F'$ as
\begin{equation}
\begin{aligned}
& X_{ia} = (X_a,C\bar X_a) \in (2,1,2,1),\\
& Y_{IA} = (Y_A,C\bar Y_A) \in (1,2,1,2),\\
& \xi_{iA} = (\xi_A,C\bar\xi_A) \in (2,1,1,2),\\
& \eta_{Ia} = (\eta_a,C\bar\eta_a) \in (1,2,2,1).
\end{aligned}
\end{equation}
Note that the scalar potential $V$ would be invariant under the $SU(2)^4$ if we naively ignore
the distinct charges of $X_a, Y_a$ and $\bar X_a, \bar Y_a$. It is not clear to us how to see the $SU(2)^4$ symmetry
in the fermion-boson coupling, due to the difficulty of describing the 't Hooft operator in the Lagrangian formalism.

Instead, let us compare the chiral primaries and $SU(2)_F\times SU(2)_F'$ in SQED-2 and model II${}_2$. Let
$(Q_i,\tilde Q^i)$, $i=1,2$, be the complex scalars in the hypermultiplets of SQED-2. In ${\cal N}=2$ language, we have
the chiral primary operators $Q_1\tilde Q^2$, $Q_2\tilde Q^1$ and $Q_1\tilde Q^1-Q_2\tilde Q^2$, in a triplet of $SU(2)_F$.
The ${\cal N}=4$ R-symmetry completes them into a multiplet $(3,3,1)$ under $SU(2)_F\times SU(2)_L\times SU(2)_R$.
In model II${}_{2}$, there is the chiral primary $X\tilde X$, completely by R-symmetry into the triplet
\begin{equation}\label{xxpr}
X\tilde X,~~~\bar X\bar{\tilde X},~~~|X|^2-|\tilde X|^2
\end{equation}
in the representation $(3,1)$ of $SU(2)_L\times SU(2)_R$. With the 't Hooft operators that correspond to
$(+1,-1)$ or $(-1,+1)$ units of magnetic flux on the $S^2$ (denoted by $C$ and $C^{-1}$), we also have
\begin{equation}
C^{-1}X^2,~~~C^{-1}\bar{\tilde X}^2, ~~~C^{-1}X\bar{\tilde X},~~~
C\bar X^2,~~~C\tilde X^2,~~~C\bar X\tilde X,
\end{equation}
which are protected to have dimension 1. $C^{-1}X^2$ for instance, under the state/operator mapping,
corresponds to two $X$ particles in their ground state
on the $S^2$ with $(-1,+1)$ units of magnetic fluxes. Together with (\ref{xxpr}), they form a multiplet in
the representation $(3,1,3,1)$ of $SU(2)_F\times SU(2)_F'\times SU(2)_L\times SU(2)_R$.

Similarly, we can identify the $SU(2)_F$ currents as
\begin{equation}
\bar X_a \overleftrightarrow{D}_\mu X^a + \bar\xi_A \sigma_\mu\xi^A,~~~
C^{-1} X_a \overleftrightarrow{D}_\mu X^a + C^{-1}\xi_A \sigma_\mu\xi^A,~~~
C \bar X_a \overleftrightarrow{D}_\mu \bar X^a + C\bar \xi_A \sigma_\mu\bar \xi^A.
\end{equation}
They are in the same supermultiplet as the dimension 1 operators above.

\subsection{Model IV and ${\cal N}=4$ QED with $N_f=3$}

Let us consider ${\cal N}=4$ $U(1)$ gauge theory with $N_f=3$
hypermultiplet matters. It can be engineered by suspending a
D3-brane between two NS5-branes, and intersecting three D5-branes
in between. The 5-branes are separated along $x^3$ direction as
before, with the NS5-branes at $x^3=0,L$, and the D5-branes at
$x^3=y_1,y_2,y_3$, with $0<y_1<y_2<y_3<L$. Now let us move the
D5-branes at $y_1$ and $y_3$ to the left and right of the two
NS5-branes, i.e. $y_1<0<y_2<L<y_3$, again inducing the creation of
stretched D3-branes. The D3-brane is then suspended from the
D5-brane at $x^3=y_1$ to the D5-brane at $y_3$. Further performing
a $\tau\to \tau/(1-\tau)$ turns the D5-branes into $(1,1)$
5-branes. The low energy world volume theory on the suspended
D3-brane is now the ${\cal N}=4$ $U(1)_{-1}\times U(1)_1\times
U(1)_{-1}\times U(1)_1$ CSM theory with three hypermultiplets
$(X,\tilde X)$, $(Y,\tilde Y)$, $(Z,\tilde Z)$, of charges
$(+1,-1,0,0)$, $(0,+1,-1,0)$ and $(0,0,+1,-1)$, which we called
model IV at $k=1$.

We can consider the more general brane configuration, with the
$(1,1)$ 5-branes replaced by $(1,k)$ 5-branes, at the suitable
angles to preserve ${\cal N}=3$ supersymmetry, which is enhanced
to ${\cal N}=4$ when the axio-dilaton lies on a particular curve.
The infrared theory does not depend on the choice of $\tau$, and
hence also possesses this ${\cal N}=4$ supersymmetry. After
T-duality and lifting to M-theory, we obtain a fractional M2-brane
at the origin of the toric hyperk\"ahler orbifold $X_8'=(({\mathbb
C}^2/{\mathbb Z}_2)\times ({\mathbb C}^2/{\mathbb Z}_3))/{\mathbb
Z}_k$. The low energy world volume theory is model IV${}_k$. The
${\cal N}=4$ supersymmetry is now evident from the $SU(2)_L\times
SU(2)_R$ symmetry of $X_8'$ for general $k$. In ${\cal N}=2$
language, the superpotential is
\begin{equation}
W = {4\pi \over k} (X\tilde X-Z\tilde Z)Y\tilde Y
\end{equation}
By examining the F-flatness and D-flatness conditions, we again find two branches of the Higgs moduli space, ${\cal M}_{H_1}$ parameterized by $(Y,\tilde Y)$, with $X=\tilde X=Z=\tilde Z=0$, and ${\cal M}_{H_2}$ parameterized by $(X,\tilde X,Z,\tilde Z)$ unconstrained, with $Y=\tilde Y=0$.

The SQED-3 SCFT on the other hand, has Coulomb branch moduli space ${\cal M}^{SQED}_C\simeq {\mathbb C}^2/{\mathbb Z}_3$, and Higgs branch moduli space ${\cal M}^{SQED}_H$ being the hyperk\"ahler quotient
\begin{equation}
\left\{\sum_{i=1}^3 (|Q_i|^2-|\tilde Q^i|^2)=0,~~
\sum_{i=1}^3 Q_i \tilde Q^i=0\right\}/U(1)
\end{equation}
which has $SU(3)$ isometry.

The classical moduli space of model IV${}_{k=1}$ has two branches, isomorphic to ${\mathbb C}^2$ and ${\mathbb C}^4$, meeting at the origin. However, as we have argued in section 3, when the massive hypermultiplets $X$ and $Z$ are integrated out, the branch ${\cal M}_{H_1}$ is corrected into ${\mathbb C}^2/{\mathbb Z}_3$.
This is also suggested by the fractional M2-brane picture.
Furthermore, ${\cal M}_{H_2}$ is singular along the locus
\begin{equation}
{\cal S}:~~|X|^2-|\tilde X|^2-|Z|^2+|\tilde Z|^2=X\tilde X-Z\tilde Z=0,
\end{equation}
where $(Y,\tilde Y)$ become massless. The effect of integrating out $(Y,\tilde Y)$ is to turn ${\cal M}_{H_2}$ into
\begin{equation}\label{glo}
\left\{|X|^2-|\tilde X|^2-|Z|^2+|\tilde Z|^2=|Q|^2-|\tilde Q|^2,~~ X\tilde X - Z\tilde Z = Q\tilde Q\right\}/U(1)
\end{equation}
where the $U(1)$ acts on $(X,\tilde X, Z, \tilde Z, Q,\tilde Q)$ with charges $(1,-1,-1,1,-1,1)$.
So we see that the quantum corrected moduli spaces ${\cal M}_{H_1}$
and ${\cal M}_{H_2}$ of the CSM theory at $k=1$ are precisely the same as
${\cal M}^{SQED}_C$ and ${\cal M}^{SQED}_H$!
We conjecture that model II$(3)_2$ (or model IV$_1$) is the same as the IR SCFT of ${\cal N}=4$ QED with $N_f=3$.

\subsection{$N_f>3$}

The IR SCFT of ${\cal N}=4$ QED with $N_f$ flavors has Coulomb branch moduli space ${\cal M}_C^{SQED}={\mathbb C}^2/{\mathbb Z}_{N_f}$ and Higgs branch moduli space ${\cal M}_H^{SQED}$ given by the hyperk\"ahler quotient
\begin{equation}
\left\{\sum_{i=1}^{N_f} (|Q_i|^2-|\tilde Q^i|^2)=0,~~
\sum_{i=1}^{N_f} Q_i \tilde Q^i=0\right\}/U(1)
\end{equation}
${\cal M}_H^{SQED}$ is the singular limit of $T^*{\mathbb {CP}}^{N_f-1}$ where the ${\mathbb{CP}}^{N_f-1}$
is shrunk to zero size. Once again, these are the same as the two branches of the quantum corrected moduli space ${\cal M}_Y$ and ${\cal M}_X$ of model II$(N_f)_{k=2}$.
This leads us to conjecture that the ${\cal N}=4$ SCFT described by model II$(N_f)_{k=2}$ is the same as SQED-$N_f$. Although, unlike the $N_f=2,3$ cases, we do not know a brane construction that motivates this identification.

\subsection{A simple nonabelian generalization}

The 2+1 dimensional ${\cal N}=4$ $U(N)$ SQCD with $N_f$ flavors can be engineered by suspending $N$ D3-branes
between a pair of parallel NS5-branes, intersecting $N_f$ D5-branes, as follows

\begin{tabular}{ccccccccccc}
& 0& 1& 2& 3& 4& 5& 6& 7& 8& 9 \\
NS5& $\times$& $\times$& $\times$& 0& $\times$ & $\times$& $\times$ & & &\\
NS5$'$& $\times$&$\times$ & $\times$ & $L$& $\times$ & $\times$& $\times$ & & &\\
D5${}_i$ & $\times$& $\times$ & $\times$ & $y_i$&  & & &$\times$ & $\times$& $\times$ \\
D3 & $\times$ & $\times$ & $\times$ & $\times$ &
\end{tabular}

\noindent where $0<y_i<L$, $i=1,\cdots,N_f$. For $N_f\leq 3$, we
can move some of the D5-branes to the outside of the pair of
NS5-branes, leaving only one D5-brane in between the NS5-branes.
In doing this an additional D3-brane is created stretching between
an NS5-brane and the D5-brane on the outside \cite{Hanany:1996ie}.
We can then perform the $SL(2,{\mathbb Z})$ duality $\tau\to
\tau/(1-\tau)$ and turn the D5-branes into $(1,1)$ 5-branes as
before, and obtain in the low energy limit ${\cal N}=4$ CSM
theories with gauge groups and representation of hypermultiplet
matter fields
\begin{equation}
\begin{aligned}
&(1)~~~ U(N)_1\times U(N)_{-1},~~~~({\bf N},{\bf\overline N}); \\
&(2)~~~ U(N)_1\times U(N)_{-1}\times U(1)_1,~~~~({\bf N},{\bf\overline N},0)\oplus ({\bf 1},{\bf N},-1); \\
&(3)~~~ U(1)_{-1}\times U(N)_1\times U(N)_{-1}\times U(1)_1,~~~~(+1,{\bf \overline N},{\bf 1},0)\oplus(0,{\bf N},{\bf\overline N},0)\oplus (0,{\bf 1},{\bf N},-1).
\end{aligned}
\end{equation}
Among these, theory (1) is the simplest example of ${\cal N}=4$ CSM theory studied by \cite{Gaiotto:2008sd},
at level $k=1$. It is natural to propose that these CSM theories describe the IR SCFT of ${\cal N}=4$ $U(N)$ QCD with
$N_f=1,2,3$ respectively.

\section{${\cal N}=2$ Chern-Simons-matter theories as quantum critical points}

\subsection{Model I: the classical theory}

Let us now consider ${\cal N}=2$ Chern-Simons-matter theory
with gauge group $U(1)\times U(1)$, at Chern-Simons level
$(k,-k)$. The fields in the gauge multiplet are denoted as
$(A_\mu,\sigma,D;\chi)$ and $(\tilde A_\mu, \tilde\sigma, \tilde
D; \tilde\chi)$ as before. The matter fields are taken to be $M_1$ chiral multiplets
$X_i$ ($i=1,\cdots,M_1$) with charge $(+1,+1)$, and $M_2$ chiral
multiplets $Y_{i'}$ ($i'=1,\cdots,M_2$) with charge $(+1,-1)$. The
scalar potential before integrating out the auxiliary fields is
\begin{equation}
\begin{aligned}
& {k\over2\pi} D\sigma - {k\over2\pi} \tilde D\tilde\sigma + \sum_i|(\sigma+\tilde\sigma)X_i|^2 +
\sum_{i'} |(\sigma-\tilde\sigma)Y_{i'}|^2 \\
& + D(\sum_i |X_i|^2+\sum_{i'}|Y_{i'}|^2) +\tilde D (\sum_i |X_i|^2-\sum_{i'} |Y_{i'}|^2)
\end{aligned}
\end{equation}
Integrating out $D$ and $\tilde D$ sets
\begin{equation}
\begin{aligned}
& \sigma = -{2\pi\over k} (\sum_i |X_i|^2+\sum_{i'}|Y_{i'}|^2), \\
&\tilde\sigma = {2\pi\over k} (\sum_i |X_i|^2-\sum_{i'}|Y_{i'}|^2).
\end{aligned}
\end{equation}
So we obtain the scalar potential
\begin{equation}
V = {16\pi^2\over k^2} \left[ \left(\sum_i |X_i|^2\right)^2 \left(\sum_{i'} |Y_{i'}|^2\right)
+\left(\sum_{i'} |Y_{i'}|^2\right)^2 \left(\sum_{i} |X_{i}|^2\right) \right]
\end{equation}
In the case $M_1=M_2=2$, this is exactly the same as the scalar potential of model II${}_k$,
but the fermion couplings will be different.
We see that there are two branches of the Higgs branch moduli
space, parameterized by
\begin{equation}
{\cal M}_{X}:~~~X_i\not=0,~~~Y_{i'}=0,
\end{equation}
and
\begin{equation}
{\cal M}_{Y}:~~~Y_{i'}\not=0,~~~X_{i}=0.
\end{equation}
On each of these Higgs branches, one combination of the $U(1)
\times U(1)$ gauge symmetry acts trivially. The abelian
Chern-Simons action can be written as
\begin{equation}\label{twistact} S_{CS} = \frac{k}{4\pi} \int A
\wedge dA - \tilde A \wedge d \tilde A = \frac{k}{4\pi} \int
a\wedge d b,\end{equation} where $a = A +\tilde A$, $b=A-\tilde
A$. Dirac quantization condition implies that the fluxes of $a$
and $b$ are independently quantized to be integer valued. Let us
assume that $k$ is even, so it follows that the constant gauge
transformations of each $U(1)$ are the subgroup ${\mathbb
Z}_{k/2}$. 
The
classical moduli space is then
\begin{equation}
\begin{aligned}
&{\cal M}_{X}\simeq {\mathbb C}^{M_1}/{\mathbb Z}_{k/2},~~
{\cal M}_{Y}\simeq {\mathbb C}^{M_2}/{\mathbb Z}_{k/2},~~k~{\rm even}
\end{aligned}
\end{equation}
The CSM theory has $SU(M_1)\times SU(M_2)$ flavor symmetry. $SU(M_2)$ acts trivially on ${\cal M}_X$, and $SU(M_1)$ acts trivially on ${\cal M}_Y$. This indicates that quantum effects cannot join the two branches of the moduli space. However, each branch may be deformed or lifted entirely.
The Chern-Simons-matter SCFT lives at the singular origin where the two branches meet.

The fermion-boson coupling in this theory is given by
\begin{equation}
\begin{aligned}
L_F &=-(\sigma+\tilde\sigma) \sum_i \bar\xi_i\xi_i-(\sigma-\tilde\sigma) \sum_{i'}\bar\eta_{i'}\eta_i
+{2\pi\over k} (\sum_i \bar X_i \xi_i+\sum_{i'}\bar Y_{i'}\eta_{i'})^\dagger
(\sum_i \bar X_i \xi_i+\sum_{i'}\bar Y_{i'}\eta_{i'})\\
&~~~
-{2\pi\over k}(\sum_i \bar X_i \xi_i-\sum_{i'}\bar Y_{i'}\eta_{i'})^\dagger
(\sum_i \bar X_i \xi_i-\sum_{i'}\bar Y_{i'}\eta_{i'})
\\
&= {4\pi\over k} \left[ \sum_i|X_i|^2 \sum_{i'}\bar\eta_{i'}\eta_{i'}
+ \sum_{i'}|Y_{i'}|^2 \sum_i \bar \xi_i \xi_i
+\sum_{i'}Y_{i'}\bar\eta_{i'} \sum_i \bar X_i \xi_i
+\sum_i X_i \bar\xi_i \sum_{i'} \bar Y_{i'} \eta_{i'}\right]
\end{aligned}
\end{equation}
We will refer to this CSM theory as ``Model I".
A bosonic version of this model describing the critical point of a triangular lattice antiferromagnetic was
studied recently in \cite{CMT}.

\subsection{A nonabelian version of model I}

Here we briefly discuss a nonabelian generalization of model I,
which has various branches of Higgs moduli space. Consider ${\cal
N}=2$ $U(N)_k\times U(N)_{-k}$ CSM theory with $M_1$ chiral
multiplets in the representation $({\bf N},{\bf N})$ and $M_2$
chiral multiplets in $({\bf N},{\bf \overline N})$, and with
vanishing superpotential. The bosonic components of the fields
will be denoted $X_{iIA}, {Y_{i'I}}^A$, where $i,i'$ are flavor
indices, and $I,J$, $A,B$ are the gauge indices of the two
$U(N)$'s. We have the auxiliary fields
\begin{equation}
\begin{aligned}
& {\sigma_I}^J = -{2\pi\over k} \left(X_{iIA} (X^\dagger)^{iJA} + {Y_{i'I}}^A{(Y^\dagger)^{i'J}}_A\right), \\
& {\tilde\sigma}{{}_A}^B = {2\pi\over k} \left( X_{iIA} (X^\dagger)^{iIB}
- {Y_{i'I}}^B {(Y^\dagger)^{i'I}}_A \right). \\
\end{aligned}
\end{equation}
The scalar potential is
\begin{equation}
\begin{aligned}
V & = |{\sigma_I}^J X_{iJA}+{\tilde\sigma}{{}_A}^BX_{iIB}|^2 + |{\sigma_I}^J {Y_{i'J}}^A-{Y_{i'I}}^B{\tilde\sigma}{{}_B}^A|^2
\end{aligned}
\end{equation}
For simplicity let us consider the 1 flavor case, i.e. $M_1=M_2=1$. In this case we have the scalar potential
\begin{equation}
V = {16\pi^2\over k^2} \left[\sum_{I,A}\left|{Y_I}^B {(Y^\dagger)^J}_BX_{JA}+{Y_J}^B {(Y^\dagger)^J}_AX_{IB}\right|^2 +
\sum_{I,A}\left| X_{IB}(X^\dagger)^{JB}{Y_J}^A
+ X_{JB}(X^\dagger)^{JA} {Y_I}^B \right|^2
\right]
\end{equation}
We can assume that, for instance, $X$ is diagonal by $U(N)\times U(N)$ gauge symmetry. It is then clear that the classical moduli space has $N+1$ branches, given by
\begin{equation}
{\cal M}_n:~~~X = {\rm diag} (x_1,\cdots,x_n,0,\cdots,0),~~~Y=
{\rm diag} (0,\cdots,0,y_{n+1},\cdots,y_N)
\end{equation}
where $n=0,\cdots,N$. Each branch ${\cal M}_n$ is isomorphic to
${\rm Sym}^n({\mathbb C}/{\mathbb Z}_k)\times {\rm Sym}^{N-n}({\mathbb C}/{\mathbb Z}_k)$.
The different branches meet pairwise along
\begin{equation}
{\cal M}_n \cap {\cal M}_m \simeq {\rm Sym}^n({\mathbb C}/{\mathbb Z}_k)\times
{\rm Sym}^{N-m}({\mathbb C}/{\mathbb Z}_k),
~~~~(n<m)
\end{equation}
and all branches meet at the origin.

\subsection{The quantum theory}

Let us now consider quantum corrections in model I. It is well
known that integrating out massive charged multiplets will shift
the CS levels. Consequently, one may expect the moduli spaces to
be lifted. To avoid this problem, we will define the theory with
shifted ``bare" CS level. The fields $X_i$'s ($i=1,\cdots,M_1$)
have charge $+1$ with respect to $a_\mu$, and $Y_{i'}$'s
($i'=1,\cdots,M_2$) have charge $+1$ with respect to $b_\mu$,
where $a_\mu, b_\mu$ are the twisted CS gauge fields defined in
(\ref{twistact}). We will introduce bare CS levels $-M_1/2$ and
$-M_2/2$ for $a$ and $b$ respectively, and adjust the action for
the matter fields so as to continue to preserve ${\cal N}=2$
supersymmtery, so that the total CS action is written as
\begin{equation}
{k\over 4\pi}\int a\wedge db - {M_1\over 8\pi}\int a\wedge da - {M_2\over 8\pi}\int b\wedge db
\end{equation}
Classically this would lift the moduli spaces ${\cal M}_X$ and
${\cal M}_Y$. We will show that when quantum corrections are taken
into account, the moduli spaces are restored. Let us assume
nonzero values of $X_i$'s, and integrate out the massive
$Y_{i'}$'s. The mass of $Y_{i'}$ is given by
\begin{equation}
m_Y = {4\pi\over k}\sum_i |X_i|^2
\end{equation}
Integrating out $Y_{i'}$ at one loop generates the couplings
\begin{equation}\label{oneloopgen}
\int {M_2\over 8\pi m} \left[ (\partial_\mu m)^2 + (db)^2 \right]
+ {M_2\over 8\pi} \int b\wedge db
\end{equation}
The last term cancels the bare CS level $-M_2/2$ for $b_\mu$. The
kinetic term $(\partial_\mu m)^2/m$ may appear to be inconsistent
with having a K\"ahler metric on the moduli space. In fact, it has
a supersymmetric completion due to the CS coupling. We can write
the ${\cal N}=2$ supersymmetric effective action of the chiral
fields $X_i$ and vector superfields  $V_a$, and $V_b$ as
\begin{equation}\label{suplag}
\int d^4\theta \left[\sum_i\overline X_i e^{V_a} X_i -{M_1\over 8\pi}V_a \Sigma_a + {k\over 4\pi} V_a \Sigma_b
- {kM_2 \over 32\pi^2\sum_i \overline X_i e^{V_a} X_i} \Sigma_b^2\right]
\end{equation}
Here $\Sigma=iD^\alpha \bar D_\alpha V$ is the linear multiplet field strength.
In component fields, the bosonic part of the action can be written as
\begin{equation}\label{bosl}
\begin{aligned}
& {k\over 4\pi}\int a\wedge db-{M_1\over 8\pi}\int a\wedge da+ {k\over 4\pi}\int (\sigma_a D_b+\sigma_b D_a)-{M_1\over 4\pi}\sigma_a D_a
\\
&+\int |D_\mu X_i|^2+\sigma_a^2 |X|^2-D_a |X|^2
\\
&+\int {kM_2\over 32\pi^2|X|^2}\left[ -(db)^2-D_b^2+
(\partial_\mu\sigma_b)^2 - 2\sigma_a \sigma_b D_b\right] \\
&
+\int  {kM_2\over 32\pi^2|X|^4} \sigma_b^2 (|D_\mu X_i|^2-\sigma_a^2 |X|^2-D_a|X|^2)+\cdots
\end{aligned}
\end{equation}
where $|X|^2\equiv \sum_i |X_i|^2$.
In particular, we see that the equation of motion for $D_a$ implies (up to order ${\cal O}(1/k^2)$ terms)
\begin{equation}
\sigma_b ={1\over k} (M_1 \sigma_a+4\pi|X|^2) = m+{M_1\over k}\sigma_a
\end{equation}
This gives the term $M_2(\partial_\mu m)^2/m$ through ${kM_2\over 32\pi^2}|X|^{-2} (\partial_\mu\sigma_b)^2$.
Moreover, we see that arbitrary constant $X_i$ and $\sigma_a=D_a=D_b=0$ solves the equations of motion. So we recover the moduli space ${\cal M}_X$. However, the low energy theory on this moduli space is not simply a sigma model in $X_i$'s. The effective action takes the form
\begin{equation}\label{lagpre}
\begin{aligned}
& {k\over 4\pi}\int a\wedge db-{M_1\over 8\pi}\int a\wedge da
+\int |D_\mu X_i|^2 + {M_2\over 2k|X|^2} (\partial_\mu |X|^2)^2 - {kM_2\over 32\pi^2|X|^2}(db)^2
+{\rm fermions}
\end{aligned}
\end{equation}
Note that the last term in (\ref{bosl}) renormalizes the coefficient of $|D_\mu X_i|^2$ to $1-{M_2\over k}$,
but this can be absorbed by a rescaling of the field $X_i$, and the Lagrangian stays in the form (\ref{lagpre}) up to ${\cal O}(1/k^2)$ terms.
Now dualizing $b_\mu$, we have
\begin{equation}
\begin{aligned}
& \int  \tilde F_b\wedge (d\varphi+{k\over 4\pi}a) -\int  {kM_2\over 32\pi^2|X|^2}|\tilde F_b|^2
\\
&\to \int {8\pi^2 |X|^2\over kM_2} (\partial_\mu\varphi+{k\over 4\pi}a_\mu)^2
\end{aligned}
\end{equation}
We then end up with the action
\begin{equation}
\begin{aligned}
& -{M_1\over 8\pi}\int a\wedge da
+\int |D_\mu X_i|^2 + {M_2\over 2k|X|^2} (\partial_\mu |X|^2)^2 + {8\pi^2 |X|^2\over kM_2}(\partial_\mu\varphi+{k\over 4\pi}a_\mu)^2
+{\rm fermions}
\end{aligned}
\end{equation}
The $U(1)$ gauge symmetry acts as
\begin{equation}\label{uone}
X_i \to e^{i\Lambda} X_i,~~~ a_\mu\to a_\mu+\partial_\mu \Lambda,~~~\varphi\to \varphi - {k\over 4\pi}\Lambda
\end{equation}
The moduli space ${\cal M}_X$ is parameterized by $X_i$ and $\varphi$, modulo the $U(1)$ action.
In other words, ${\cal M}_X=\widetilde{\cal M}_X/U(1)$, where the metric on $\widetilde {\cal M}_X$ is
\begin{equation}
\begin{aligned}
ds^2 &= |dX_i|^2 + {M_2\over 2k|X|^2} (d|X|^2)^2 + {8\pi^2|X|^2\over k M_2}d\varphi^2
\end{aligned}
\end{equation}
This is the induced metric on $\mu^{-1}(0)\subset {\mathbb C}^{M_1+1}$, where
\begin{equation}
\begin{aligned}
&{\mathbb C}^{M_1+1}=\{(X_i,z = \sqrt{2M_2\over k}\,\rho\, e^{{2\pi i\varphi / M_2}})\},\\
& ds^2 = \sum_{i=1}^{M_1}|dX_i|^2 + |dz|^2,\\
& \mu=\sum_{i=1}^{M_1} |X_i|^2 - {k\over 2M_2}|z|^2=\sum_{i=1}^{M_1} |X_i|^2-\rho^2.
\end{aligned}
\end{equation}
and $\mu$ is the moment map of the symplectic form on ${\mathbb C}^{M_1+1}$
\begin{equation}
\begin{aligned}
\omega &= dX_i\wedge d\bar X_i + dz\wedge d\bar z \\
&=dX_i\wedge d\bar X_i+{8\pi i\over k}\rho d\rho \wedge d\varphi.
\end{aligned}
\end{equation}
with respect to the $U(1)$ action (\ref{uone}). Therefore, we have found that the one-loop corrected metric on the moduli space
${\cal M}_X$ is that of the (singular) symplectic quotient
\begin{equation}
{\cal M}_X \simeq {\mathbb C}^{M_1+1}//U(1) = \mu^{-1}(0)/U(1).
\end{equation}
The $U(1)$ acts on the ${\mathbb C}^{M_1+1}$ with charges $(M_2,M_2,\cdots,M_2,-{k/2})$.
Similarly, the one-loop correction turns ${\cal M}_Y$ into the symplectic quotient of ${\mathbb C}^{M_2+1}$ by the $U(1)$
acting with charges $(M_1,M_1,\cdots,M_1,-{k/2})$. Note that the massless fields along the moduli space
still couple nontrivially to a $U(1)$ Chern-Simons gauge field, at level $-M_1/2$ for ${\cal M}_X$
and level $-M_2/2$ for ${\cal M}_Y$.

If the bare CS level for $a_\mu$ is not $-M_1/2$, the moduli space
${\cal M}_X$ still exists, whereas the moduli space ${\cal M}_Y$
will be lifted. Similarly, if the CS level for $b_\mu$ is not
$-M_2/2$, the moduli space ${\cal M}_X$ would be lifted. This is
because in (\ref{suplag}) there would be an additional
supersymmetric CS term
\begin{equation}
{k_b\over 4\pi} \int d^3x \int d^4\theta V_b \Sigma_b = {k_b\over 4\pi} \int \left(b\wedge db +2 \sigma_b D_b + \bar\chi_b \chi_b\right)
\end{equation}
As a consequence, $\sigma_a=D_a=D_b=0$ would not be a solution to the equation of motion for nonzero constant $X_i$. In fact, the moduli space of $X_i$ will be lifted by a potential
\begin{equation}
V\sim {k_b^2\over k^4}(|X|^2)^3
\end{equation}
This would be a two-loop contribution to the effective Lagrangian, which is why it was absent from (\ref{oneloopgen}) but is needed for the supersymmetric completion of the one-loop effective Lagrangian.

\section{Summary and outlook}

We have presented examples of abelian CSM theories with
${\cal N}=4$ supersymmetries that describe quantum critical
points where different branches of the moduli space meet. A new feature of the
hypermultiplet moduli spaces in CSM theories is that they can receive quantum corrections,
but can still be determined exactly due
to the rigidity of the hyperk\"ahler metric. These
theories can also be studied perturbatively at weak coupling in $1/k$.

We then considered ${\cal N}=2$ abelian and nonabelian CSM theories.
It is easy to describe classical theories with multiple Higgs branches. Quantum mechanically,
the moduli space receives nontrivial corrections and may get lifted. We have analyzed the abelian case,
and described ${\cal N}=2$ CSM theories with two Higgs branches that are lifted in the classical theory but are restored in the
quantum theory. We computed the one-loop correction to the moduli space and found
that each branch is turned into a symplectic quotient of the form ${\mathbb C}^{M+1}//U(1)$.
The moduli spaces of the nonabelian theories are more complicated.
We hope to explore them in future works. It would be particularly interesting
to find an example of CSM theory describing a quantum critical
point, with a nontrivial 't Hooft limit that also has a brane
construction which allows one to identify its gravity dual.

We proposed that the model
III at $k=1$ (or model II$_{k=2}$) and model IV at $k=1$ (or model II$(3)_{k=2}$) describe the IR SCFT of
${\cal N}=4$ QED with $N_f=2$ and $N_f=3$, respectively. The quantum corrections
to the moduli spaces are crucial for the identification,
and provide highly nontrivial checks of the conjecture.
Despite
 that such CSM theories are strong coupled, they belong to
a family of CSM theories parameterized by the CS level $k$. One
could hope to learn aspects of the strongly coupled theories
by extrapolating from the weakly coupled ones, i.e. at large $k$. The
mechanism of enhanced global symmetries at small $k$ through 't
Hooft operators clearly deserves further study. We have further proposed
CSM descriptions of ${\cal N}=4$ QED with
$N_f>3$ based on matching the quantum moduli spaces. It would be nice to have
brane constructions for this identification.

We also generalized such constructions to
${\cal N}=4$ $U(N)$ SQCD with $N_f$ flavors, for $N_f=1,2,3$. Since the $U(1)$ gauge group does not decouple in SQCD, one would really
like to understand the ``genuinely nonabelian''  ${\cal N}=4$
$SU(N)$ SQCD, say for $N=2$, and $N_f$ flavors. It is conceivable
that there are CSM descriptions of the IR SCFTs of these theories
as well (in particular, they should reproduce the Coulomb and Higgs branch moduli spaces
of SQCD), although we do not have any proposals so far.

\subsection*{Acknowledgments}

We are grateful to Davide Gaiotto, Subir Sachdev, Matt Strassler
and Cenke Xu for discussions. We would like to thank the
organizers of the workshop {\sl AdS/CFT, Condensed Matter and QCD}
at McGill University for their hospitality during the final stage
of this work. The work of DLJ is supported by  DOE grant
DE-FG02-96ER40959. The work of XY is supported by the Center for
the Fundamental Laws of Nature at Harvard University.

\end{document}